# Suppressed effective viscosity in the bulk intergalactic plasma

I. Zhuravleva[1,2,3], E. Churazov[4,5], A. A. Schekochihin[6,7,8], S. W. Allen[2,3,9], A. Vikhlinin[10] & N. Werner[11,12,13]

[1]Department of Astronomy & Astrophysics, University of Chicago, 5640 S Ellis Ave, Chicago, IL 60637, USA

[2]KIPAC, Stanford University, 452 Lomita Mall, Stanford, CA 94305, USA

[3]Department of Physics, Stanford University, 382 Via Pueblo Mall, Stanford, CA  94305-4060, USA

[4]Max Planck Institute for Astrophysics, Karl-Schwarzschild-Strasse 1, D-85741 Garching, Germany

[5]Space Research Institute (IKI), Profsoyuznaya 84/32, Moscow 117997, Russia

[6]The Rudolf Peierls Centre for Theoretical Physics, University of Oxford, Clarendon Laboratory, Parks Road, Oxford OX1 3PU, UK

[7]Merton College, Oxford OX1 4JD, UK

[8]Niels Bohr International Academy, Blegdamsvej 17, DK-2100, Copenhagen, Denmark

[9]SLAC National Accelerator Laboratory, 2575 Sand Hill Road, Menlo Park, CA 94025, USA

[10]Harvard-Smithsonian Center for Astrophysics, 60 Garden Street, Cambridge, MA 02138, USA

[11] MTA-Eötvös University Lendulet Hot Universe Research Group, Pázmány Péter sétány 1/A, Budapest, 1117, Hungary

[12]Department of Theoretical Physics and Astrophysics, Faculty of Science, Masaryk University, Kolářská 2, Brno, 611 37, Czech Republic

[13]School of Science, Hiroshima University, 1-3-1 Kagamiyama, Higashi-Hiroshima 739-8526, Japan

**Transport properties, such as viscosity and thermal conduction, of the hot intergalactic plasma in clusters of galaxies, are largely unknown. While for laboratory plasmas these characteristics are derived from the gas density and temperature[1], such recipes can be fundamentally different for the intergalactic plasma[2] due to a low rate of particle collisions and a weak magnetic field[3]. In numerical simulations, one often cuts through these unknowns by modeling these plasmas as hydrodynamic fluids[4-6], even though local, non-hydrodynamic features observed in clusters contradict this assumption[7-9]. Using deep Chandra observations of the Coma Cluster[10-11], we probe gas fluctuations in intergalactic medium down to spatial scales where the transport processes should prominently manifest themselves - at least if hydrodynamic models[12] with pure Coulomb collision rates were indeed adequate. We find that they do not, implying that the effective isotropic viscosity is orders of magnitude smaller than naively expected. This indicates an enhanced collision rate in the plasma due to particle scattering off microfluctuations caused by plasma instabilities[2,13,14], or that the transport processes are anisotropic with respect to local magnetic field[15]. For that reason, numerical models with high Reynolds number appear more consistent with observations. Our results also demonstrate that observations of turbulence in clusters[16,17] are becoming a branch of astrophysics that can sharpen theoretical views on such plasmas.**

The Coma Cluster is a nearby, massive, hot (mean gas temperature ~ 8 keV, Supplementary Figure 1) galaxy cluster that has undergone several recent mergers with intermediate-mass subgroups[18]. X-ray observations show that the gas in this cluster is disturbed (Figure 1) and is likely turbulent[10,19]. Radio observations of Coma reveal extended non-thermal emission, suggesting the presence of a magnetic field with the strength ~ several μG[20], which gives the ratio of thermal to magnetic pressure ~ 100. At the same time, the plasma is weakly collisional, i.e., the time between particles' collisions via Coulomb interaction is longer than their gyration period in the magnetic fields. If such a plasma were described by a basic model of hydrodynamic fluid with a standard set of transport coefficients determined by Coulomb collisions, its turbulent motions would be strongly suppressed on spatial scales comparable to the Coulomb mean free path, λ, of electrons and protons. For the hot, nearby Coma Cluster such scales become accessible if one observes a region offset a few hundreds kpc away from the center. In this region, the lower density increases λ up to ~ 30 kpc (Supplementary Figure 1), which can be probed by modern X-ray observatories, given sufficiently long observations. We have carried out such an exceptional, almost 12-day-long observation of the Coma offset region with Chandra. The observation delivered a deep, high-resolution image shown in Figure 1. The data reduction was done following the standard data-analysis procedure and employing the Chandra calibration files from CalDB v.4.7.8. Dividing this cluster image by the best-fitting model and applying the Δ-variance method[21], we calculated the power spectrum of surface-brightness fluctuations in this region and deduced the amplitude of density fluctuations[10,16] (Figure 2). This amplitude is ~ 4% on the smallest spatial scale that we reliably probe, ~ 35 kpc. This scale is comparable to the Coulomb mean free path in this region.

While direct measurements of the velocity field must await the next generation of X-ray telescopes, the observed density fluctuations can be used as a proxy for the turbulent velocity field in stratified atmospheres of galaxy clusters, provided that the gas motions are subsonic and driven on large scales[22,23]. In this case, at any given spatial scale l, characterized by a wavenumber k=1/l, there should be an approximately linear relation between the magnitudes of velocity and density fluctuations, $A_{3D}(k)=(4\pi P(k)k^3)^{0.5} \propto V_{1k}/c_s$, where P(k) is the power spectrum of density fluctuations, $c_s$ is the sound speed and $V_{1k}$ is the velocity along one direction. Figure 2 (also, Supplementary Figure 2) shows the velocity amplitude (normalized by the sound speed) inferred from the density fluctuations amplitude in Coma. From the lack of the amplitude steepening at scales approaching λ, it immediately follows that the effective viscosity (i.e., the effective spectral cutoff scale) is suppressed in the bulk intergalactic gas, in contrast with expectations for hydrodynamic gas with purely Coulomb collisions. Note that in the central, ~ 250 kpc region of the cluster we probe the amplitude of density fluctuations on similar or even smaller scales (Supplementary Figure 2), however, λ is about an order of magnitude lower in this region.

It is useful to illustrate the above qualitative conclusion by comparing the observed spectra with the results of Direct Numerical Simulations (DNS) of incompressible hydrodynamic turbulence, which numerically solve the Navier-Stokes equation. DNS provide power spectra of turbulent velocities and of so-called passive scalars, i.e., fluid characteristics that are advected by gas velocities but do not affect the motions[12,24]. We have argued earlier[22] that at sufficiently small scales, density fluctuations in galaxy clusters behave approximately as a passive scalar. Therefore, the scalar spectra predicted by DNS should be relevant for this comparison. The shapes of the DNS scalar spectra at scales well below the driving scale are fully specified by a combination of two numbers: the Kolmogorov microscale and the diffusivity of the passive scalar. The Kolmogorov microscale is defined as $\eta=(\nu_0^3/\varepsilon)^{1/4}$, where $\nu_0$ is the kinematic gas viscosity that is proportional to the mean free path and the proton thermal speed and can be evaluated directly from the gas density and temperature for a putative plasma dominated by Coulomb collisions[25], and ε is the rate of turbulent dissipation in Coma, which we estimated from the observed amplitude of density fluctuations (Figure 2), assuming that the latter follows the Kolmogorov scaling as $\varepsilon=C_Q V_{1k}^3 k=C_Q(A_{3D} c_s)^3 k$, where $C_Q=3^{3/2} 2\pi/(2C_k)^{3/2} \cong 5$ [16] (note that turbulent velocities in the Perseus cluster estimated on this basis are consistent with those measured directly by the Hitomi satellite[17,26]). For our estimates, we used the measured amplitude, $A_{3D}$, at scales ~ 200 kpc. With these definitions, the Kolmogorov microscale is ~ 20 (~ 50) kpc in the central (offset) region (Supplementary Figure 1). While the gas velocities are subject to the viscosity of the fluid (momentum diffusivity), the passive scalar has its own diffusivity. For subsonic motions, the pressure remains smooth, and, therefore, any density fluctuations should be compensated by anticorrelated temperature fluctuations (this local pressure balance is established on a time scale faster than that of the turbulent motions). These temperature fluctuations and, consequently, the density fluctuations, are subject to thermal conduction. The strength of the latter is parameterized by the thermal Prandtl number, Pr, defined as the ratio of kinematic viscosity to thermal conductivity. For a plasma dominated by Coulomb collisions, the Prandtl number is simply the square root of the electron to ion mass ratio, which is ~ 0.02. This implies that the temperature fluctuations (and density fluctuations) will be damped faster, i.e., on larger scales, than the velocities. This argument has already been used to constrain the level of conductivity in the Coma Cluster[23]. However, the suppression of electron thermal conductivity[27,28] could increase Pr much above 0.02. Therefore, rather than trying to select the most plausible value of the Prandtl number, we simply interpret Pr as an additional parameter associated with DNS spectra.

Figure 2 shows the comparison between simulated (DNS) and observed spectra. It is clear that the assumption of a Coulomb-collision-dominated plasma is ruled out for the Coma Cluster: the slopes of the observed amplitudes are significantly shallower than those formed in simulations even after the limitations of the $\Delta$-variance method[21] are taken into account. This is true for Pr=1 and Pr < 1 makes this discrepancy even stronger (the Pr > 1 regime is physically unlikely in the ICM and we discuss it separately in Methods and Supplementary Figure 3 for completeness).

This conclusion is robust against any reasonable uncertainty in the estimates of $\eta$ and $\varepsilon$ (Methods). It is interesting to note that the comparison shows that if the effective viscosity in Coma were comparable to the Spitzer viscosity, it would be affecting gas properties on relatively large spatial scales, up to ~ 400 kpc, which are comparable with the typical injection scales in cosmological simulations of galaxy clusters.

The suppressed gas viscosity in the intergalactic plasma could be a consequence of the interaction of particles and plasma instabilities. Electromagnetic fluctuations induced by plasma instabilities could be present at scales much smaller than $\lambda$, down to the ion gyroscale and below. Interactions between fluctuations and particles may increase the effective collision rate and, therefore, the effective Reynolds number[2,13]. Another key effect, which is currently not well understood, is the possible modification of plasma turbulence by the anisotropy of the viscosity with respect to the local direction of the (statistically tangled) magnetic field[29]. In principle, even if the viscosity is determined by Coulomb collisions, it may be unable to damp motions that do not lead to change in the magnetic field strength. Turbulence in such a plasma has only recently started to be explored numerically[15] and it is interesting that our results may be consistent with a description of intracluster gas as a Braginskii plasma.

Let us now ask for what values of effective viscosity and diffusivity, our observations could be qualitatively consistent with the DNS? Figure 3a answers this for the case of Pr=0.1. We can rescale the DNS spectrum, varying the gas viscosity. The lower the viscosity, the farther the DNS spectra shift towards high k. For effective viscosity ~ 0.01 (~ 0.05) of the Spitzer value, the DNS data for Pr=0.1 becomes qualitatively consistent with the data in the central (offset) region (Figure 3a). Following the same logic for other values of Pr, we found the constraints on the minimal effective viscosity in Coma shown in Figure 3b. In conclusion, in order to have consistency between the observations and DNS simulations, the effective viscosity must be suppressed by at least a factor of ~10 to ~1000 for Pr ≤ 1.

We also compared the DNS results with the density fluctuations measured in a sample of nearby, cool-core clusters[26]. For all clusters (except Virgo and Centaurus), the core is divided into inner and outer halves. We calculated the turbulent dissipation rate using the amplitude of density fluctuations in the middle of the range of probed scales in each cluster. Kolmogorov microscale is then averaged within the considered regions. When plotted together with Coma (Figure 4), the amplitudes cover more than two orders of magnitude in the rescaled wavenumber kη. Either together or individually, these spectra cannot be described by DNS with Spitzer viscosity if default, averaged over r and k, ε and η are used. More robust conclusions require detailed consideration of individual clusters, which is beyond the scope of this paper.

As a caveat, let us stress that it is not at present possible to prove that the observed fluctuations are associated with the turbulence in clusters as we do not have direct velocity measurements. These are on the agenda for future X-ray observatories. However, we have explored how our results change if we exclude the most prominent structures associated with quasi-linear structures of low-entropy gas[11,18] and galaxies[30]. Overall, our conclusion that there is no evidence for a sharp viscous cutoff does not change (Methods, Supplementary Figure 4).

Thus, despite a degree of uncertainty (Methods, Supplementary Figures 4-6), the evidence for suppressed effective viscosity in the bulk intergalactic gas in galaxy clusters looks strong. In contrast to the expectations for a Coulomb-collision-dominated plasma, the effective Reynolds number appears to be large. From the perspective of plasma physics, this finding is consistent with the presence of plasma instabilities that, by interacting with particles, effectively increase the collision rate of plasma, and/or with the establishment of a new type of turbulence in which motions adjust to be immune to the locally anisotropic plasma viscosity. From the perspective of hydrodynamic models of galaxy clusters, our finding favors the use of high-resolution simulations with the lowest possible numerical viscosity, as opposed to physical viscosity at the Spitzer value.

**Acknowledgments** Support for this work was provided by NASA through Chandra Award Number GO6-17123X, issued by the Chandra X-ray Observatory Center, which is operated by the Smithsonian Astrophysical Observatory for and on behalf of NASA under contract NAS8-03060. E.C. acknowledges partial support by the Russian Science Foundation grant 19-12-00369. A.A.S. acknowledges partial support by grants from UK STFC and EPSRC and by the Simons Foundation via a Visiting Professorship at NBIA. NW is supported by the Lendület LP2016-11 grant awarded by the Hungarian Academy of Sciences.

**Author Contributions** I.Z.: data analysis, interpretation, manuscript preparation, principal investigator of the offset Coma Cluster observations; E.C.: data analysis, interpretation, discussions, manuscript preparation; A.A.S.: interpretation, discussions, manuscript preparation; S.W.A., A.V., N.W.: discussions and manuscript review.

**Author Information** Reprints and permissions information is available at www.nature.com/reprints. The authors declare no competing financial interests. Readers are welcome to comment on the online version of the paper. Correspondence and requests for materials should be addressed to I.Z. (zhuravleva@astro.uchicago.edu).

**Data Availability:** The observational data analyzed in this study are available in NASA's HEASARC repository (https://heasarc.gsfc.nasa.gov). The analyzed and plotted data of this study are available from the corresponding author upon reasonable request.

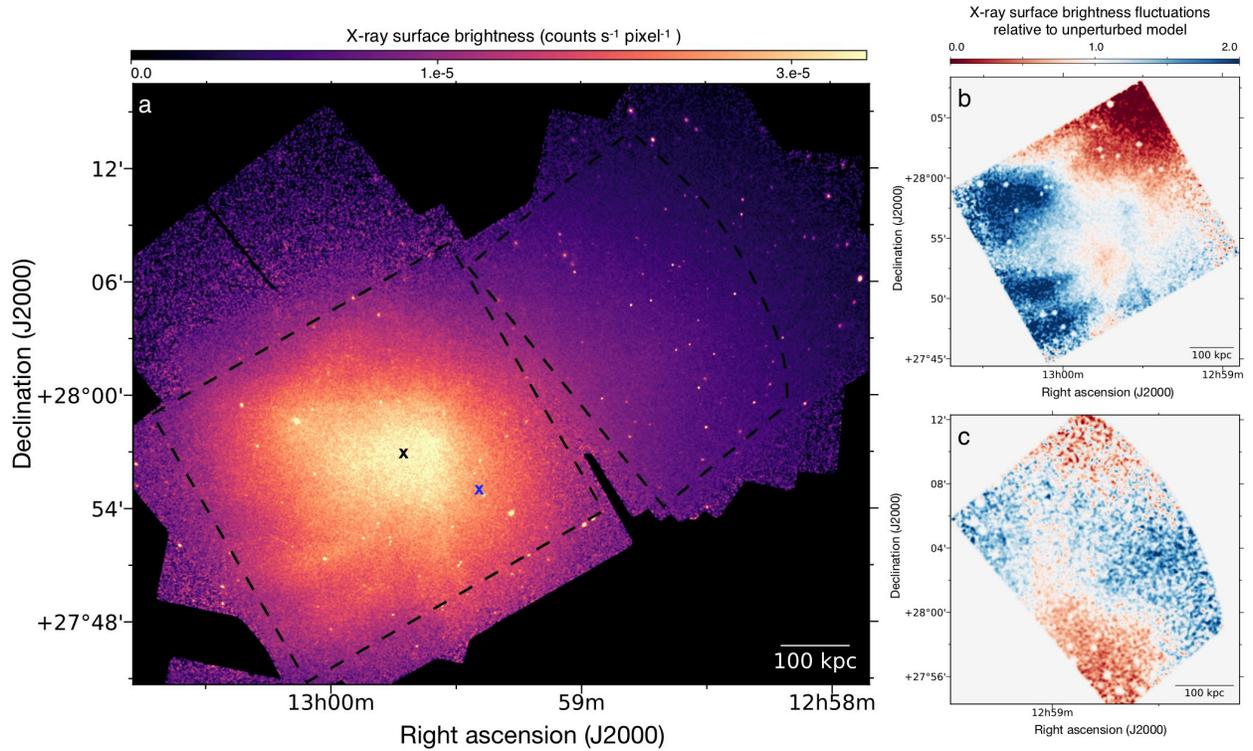

**Figure 1 | X-ray image of the Coma Cluster.** (a) X-ray surface brightness obtained in the 0.5-3 keV energy band from Chandra observations. The following ObsIDs are used: 9714, 10672, 13993 – 13996, 14406, 14410, 14411, 14415, 18271 – 18276, 18761, 18791 – 18798, 19998, 20010, 20011, 20027 – 20031, 30037 – 20039, with the total exposure ~ 1.5 Ms. Dashed regions show the central and offset regions used in the analysis. (b) Surface brightness in the central ~ 500 x 500 kpc region divided by the best-fitting model of the mean surface-brightness profile centered on RA=12h59m42.67 (J2000) and Dec=+27°56'40.9 (J2000) (black cross, see Methods and Supplementary Figure 5). (c) The same as (b) for the offset region located at the distance ~ 250-550 kpc NW from the center. In order to account for the clusters' large-scale asymmetry, the center of the model surface-brightness profile was shifted by ~ 120 kpc for this region (blue cross, see Methods and Supplementary Figure 5). White circles in panels b-c show point sources excised from the images. Images a, b, and c were lightly smoothed with a 1", 4" and 4" Gaussian for visual purposes, respectively. The redshift of the Coma Cluster is z=0.023, so that 1' corresponds to a physical scale of ~ 27.2 kpc (for h=0.72, $\Omega_m$=0.3, $\Omega_\Lambda$=0.7).

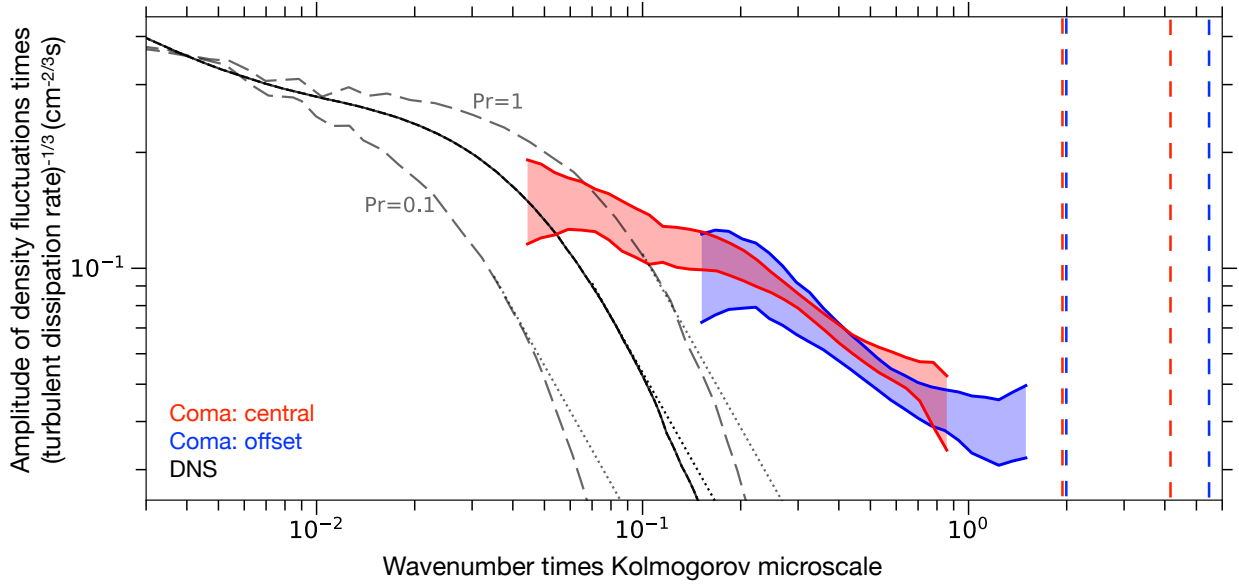

**Figure 2 | Scale-by-scale comparison of the amplitude of density fluctuations in the Coma Cluster and direct numerical simulations (DNS) of hydrodynamic turbulence.** The amplitude is shown for the central (red) and offset (blue) regions in Coma rescaled to the dissipation rate, $\varepsilon^{-1/3}$, versus the wavenumber k times the Kolmogorov microscale, $\eta$. In these units, the DNS predictions are the same for both regions. The wavenumber k is defined to be the inverse of the spatial scale l without a $2\pi$ factor. Note that when plotting the DNS spectra, we took into account a $2\pi$ factor used in the DNS definitions and used the $V_{1k}^2=(2/3)kE(k)$ relation between the velocity amplitude and the energy spectrum E(k). The width of each region shows the estimated $1\sigma$ statistical uncertainty. Black curves show the velocity[12] (solid) and passive scalar[24] (dashed) amplitudes from DNS. We show the cases of Pr=1 and 0.1 provided by the numerical simulations. In the case of Pr=0.02, the amplitude cutoff will occur at even smaller k$\eta$ as it scales with the Prandtl number as $Pr^{3/4}$. Dotted lines show how spectral slopes are modified by the $\Delta$-variance method[21] used to derive the density fluctuation amplitudes. Vertical red and blue dashed lines show the range of the Coulomb mean free path times the Kolmogorov microscales in the central and offset regions, respectively. Uncertainties in $\eta$ and $\varepsilon$ may shift theoretical spectra in the horizontal direction by a maximum factor of ~ 2 (see Methods). Other systematic uncertainties are discussed in Methods and shown in Supplementary Figures 4-6.

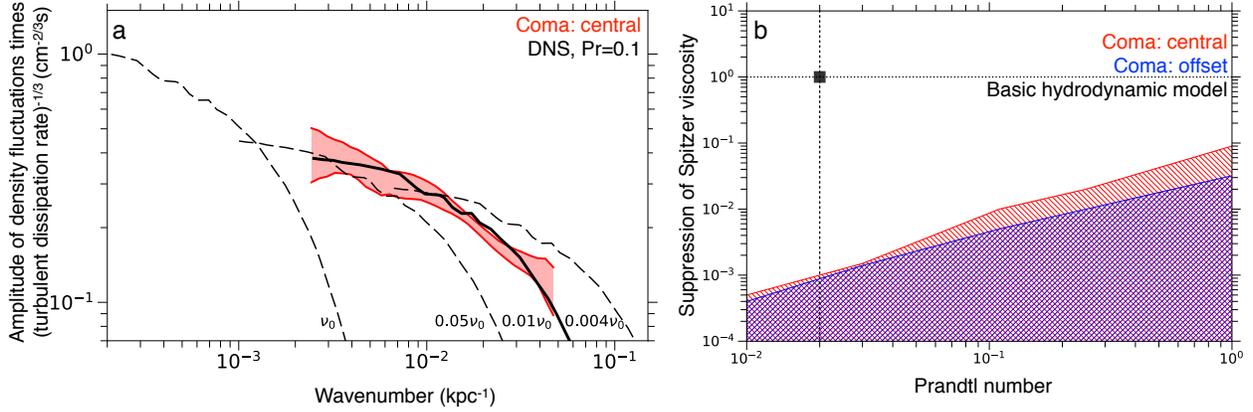

**Figure 3 | Constraints on gas viscosity in the Coma Cluster.** (a) The observed amplitude of density fluctuations in the central region in Coma (see Figure 2 for details). Black dashed curves show the passive-scalar amplitude from DNS for Pr=0.1 and the effective viscosity as a fraction of the Spitzer viscosity. The thick curve shows the DNS spectrum that most closely resembles the observations. (b) Viscosity suppression versus Prandtl number in the central (red) and offset (blue) regions in Coma. The hatched regions show the estimated values of gas viscosity and Pr that describe the observed data. The case of hydrodynamic gas with pure Coulomb collisions ($\nu = \nu_0$ and Pr=0.02) is shown with a black square in the top left corner. For Pr >1 see Methods and Supplementary Figure 3. Note that here we use default values of $\varepsilon$ and $\eta$. If their radial and scale variations are considered, the upper limits on the suppression factor may vary by a factor of few.

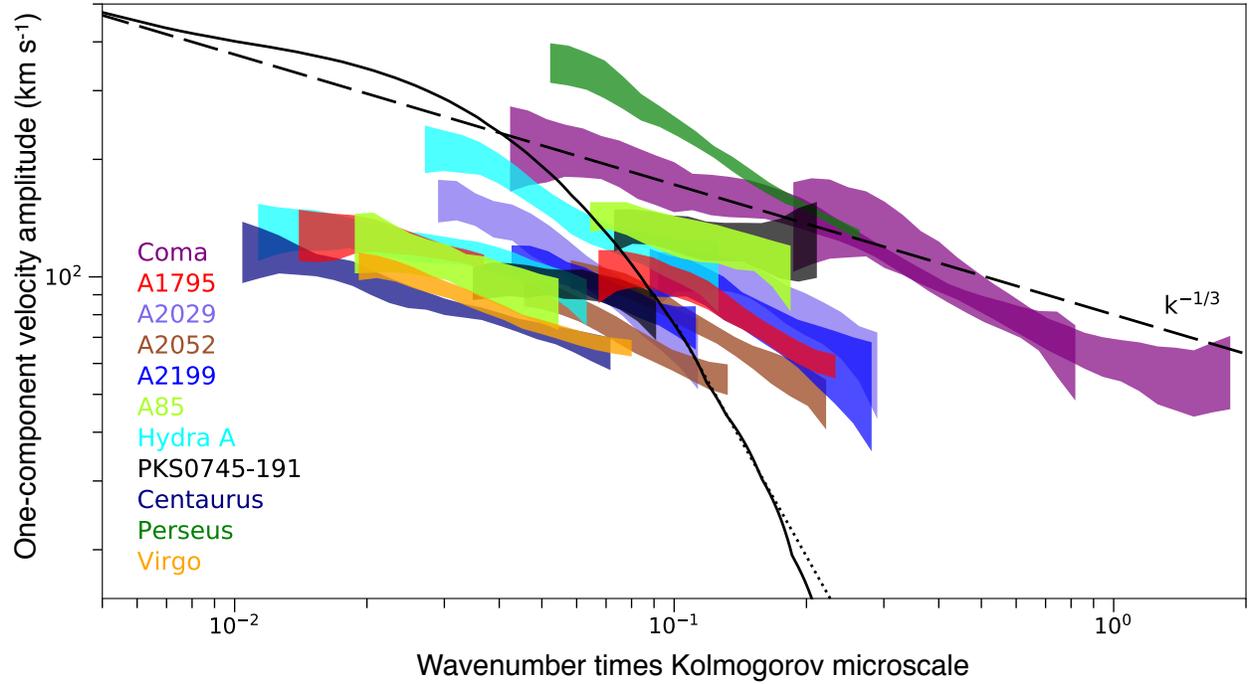

**Figure 4 | Comparison between velocity amplitudes measured in a sample of galaxy clusters and DNS simulations.** The velocity amplitudes are inferred from the amplitudes of density fluctuations and are shown as a function of wavenumber times the Kolmogorov microscale (same as in Figure 2). The width of each curve shows 1σ statistical uncertainties. Solid black curve and dotted line show the velocity amplitude from DNS and its modification by the Δ-variance method[21] used to derive the density fluctuation amplitudes. The dashed line shows the slope of Kolmogorov turbulence. For details see the Main Text.

**Methods**

**The sensitivity of the results to the variations of default parameters**

Supplementary Figure 1 shows radial profiles of the deprojected number electron density, $n_e$, electron temperature, $T_e$, and their best-fitting models in Coma. Gas in Coma is almost isothermal with $T_e \sim 8$ keV. This gives the sound speed in the gas, $c_s = (\gamma k_B T_e / \mu m_p)^{1/2} \sim 1450$ km s$^{-1}$. Here, $\gamma = 5/3$ is the adiabatic index, $k_B$ is the Boltzmann constant, $\mu = 0.61$ is the mean particle weight and $m_p$ is the proton mass. The mean free path of electrons and ions in gas dominated by Coulomb collisions is $\lambda = 23 (T_e/10^8 \text{ K})^2 (n_e/10^{-3} \text{ cm}^{-3})^{-1}$ kpc. It is shown in the panel (c) together with the kinematic viscosity $\nu_0 = \nu_{dyn}/\rho$, where $\rho$ is the mass density of the gas and $\nu_{dyn}$ is the dynamic viscosity calculated as $\nu_{dyn} = 5500$ g cm$^{-1}$ s$^{-1}$ $(T_e/10^8 \text{ K})^{5/2} (\ln\Lambda/40)^{-1}$. Here, $\Lambda$ is the Coulomb logarithm. The dissipation rate, $\varepsilon$, can be estimated from the characteristic amplitude of density fluctuations, $A_{3D}$, assuming that it follows the Kolmogorov scaling, viz., $\varepsilon = C_Q V_{1k}^3 k = C_Q (A_{3D} c_s)^3 k$. Panel (d) shows $\varepsilon$ measured from the mean value of $A_{3D}$ on spatial scales 50, 100 and 200 kpc in the central and offset regions. By default, we choose the value at 200 kpc, i.e. $\varepsilon = 0.051$ (0.019) cm$^2$s$^{-3}$ in the central (offset) region. Knowing $\varepsilon$ and $\nu_0$, we calculate the Kolmogorov microscale defined as $\eta = \nu_0^{3/4}/\varepsilon^{1/4}$. This value varies with radius and scale. By default, we choose $\eta$ at 200 kpc scale averaged within the central and offset regions, i.e. $\sim 20$ and $\sim 50$ kpc, respectively. Finally, we can measure the effective Reynolds number at each scale $l = 1/k$ as $Re = V_{1k} l / \nu_0$ (panel f).

Kinematic viscosity varies by a factor of 2.5 within the central and offset regions. The dissipation rate varies with the scale by a factor of 2-3 and by a factor of 8 if statistical uncertainties on $A_{3D}$ are taken into account. Our conservative estimates show that these variations can vary the default value of $\eta$ by a maximum factor of 2. Shifting the observed spectra shown in Figure 2 by this factor in the horizontal direction does not affect our conclusion that DNS cannot describe the data and gas viscosity is suppressed in the bulk gas. This also leads to a factor-of-few uncertainty on the upper limits of the suppression factor shown in Figure 3.

**Case of Pr > 1**

As demonstrated in the main text, the assumption of the full Spitzer viscosity and conductivity strongly contradicts the observed slope of the fluctuations spectrum for Pr ≤1. Let us describe the modification of the gas transport coefficients relative to the Spitzer plasma by coefficients $f_i$ and $f_e$, which define the change of the effective mean free path for ions and electrons, respectively (the inverse of these coefficients characterizes the change of the collision rates). With this ansatz, it is easy to show that

$$\Pr \approx \frac{\gamma}{3.93\,(\gamma-1)\mu} \frac{f_i V_i + f_e V_e \frac{m_e}{m_i}}{f_i V_i + f_e V_e}, \tag{1}$$

where $m_{e/i}$ and $V_{e/i}$ are the mass and rms velocities of electrons/ions, $\gamma=5/3$ is the adiabatic index and $\mu\sim 0.6$ is the mean particle atomic weight. The prefactor in (1) is $\sim 1$. For any values of $f_e$ and $f_i$, Pr varies between $\sim 1$ and $\sim m_e/m_i \sim 5\,10^{-4}$, i.e. Pr is always less than 1. Indeed, Pr>1 requires not only a strong suppression of the electron thermal conductivity but also the suppression of the ion thermal conductivity relative to the ion viscosity. For completeness, we consider this case below, but we note that Pr>1 by itself implies nontrivial departures from the collisional unmagnetized plasma regime.

For Pr>1, there are three distinct parts in the spectrum of a passive scalar[31]: an inertial range (for small Reynolds numbers, it may not exist at all), $E(k)\sim k^{-5/3}$ for $k<1/\eta=(\varepsilon/\nu_0^3)^{1/4}$; a diffusive range with an exponential cutoff of the scalar spectrum at $k\geq 1/\eta_B=(\varepsilon/(\nu_0\alpha^2))^{1/4}$; and a viscous-convective range, $E(k)\sim k^{-1}$, in between these two regimes. Here $\alpha$ is the thermal diffusivity of the gas and $\eta_B$ is the so-called Batchelor scale.

Given that the slopes of the observed spectra in Coma are not far from Kolmogorov's $k^{-5/3}$ law, one can expect that even in the Pr>1 case, the suppression of the viscosity will again place the Coma data into the inertial range of DNS spectra where the slope is close to 5/3. This type of solution (called type-A below) is essentially the same as already discussed in the main text and it is not sensitive to the value of the Prandtl number, as long as Pr≥1 (Supplementary Figure 3, panel c, bottom hatched region).

The modification of viscosity such that, for a given Prandtl number, the Coma spectrum falls into the viscous-convective would clearly predict a far too flat spectrum and can be excluded. This is illustrated for a specific case of Pr=5.56 in Supplementary Figure 3 (panels a-b).

Finally, theoretical spectra may resemble the observed ones if the Coma spectrum falls into the region between the viscous-convective and diffusive ranges, where the spectrum gradually changes from $k^{-1}$ to a steep cutoff. Given that the dynamic range of the observed spectra is limited, it is possible to find such a combination of viscosity and conductivity that the shape of the expected spectrum matches the observed one (Type-B solution below). Such a region is characterized by $k\eta_B \approx$ const in the DNS, and, given the definitions of the Prandtl number and $\eta_B$, requires $\nu_0^3 Pr^{-2} \approx$ const.

Using DNS for large Prandtl and modest Reynolds numbers (it is currently computationally challenging to explore the regime of both large Pr and large Re with DNS), we identified combinations of the viscosity suppression factor and Pr that broadly reproduce the observed spectrum in Coma (Supplementary Figure 3). For a given value of the viscosity suppression factor, a value of the Kolmogorov scale was estimated, and each DNS spectrum was shifted along the horizontal axis accordingly. The normalization of the DNS spectra was set to match the observed spectra in the middle of the wavenumber range. In the Figure, one can readily see both Type-A and Type-B solutions, which occupy the bottom part of the plot and a diagonal stripe, respectively. As expected in the Type-B solution (from the condition that $k\eta_B \approx$ const), the effective viscosity follows the $Pr^{2/3}$ law. For large Pr, this implies that for the Type-B solutions the viscosity has to be increased relative to Spitzer viscosity, rather than suppressed (as for type-A solutions).

Once again, we would like to emphasize that even if Pr>1 are considered here and formally provide constraints on effective viscosity that make it possible to match the observed and numerical spectra, such Pr are not possible in a plasma where collision rates are the only parameters that affect the transport. Also, the limited dynamic range of the observed spectra precludes more stringent limits on the effective viscosity. Thus, on the basis of available evidence and theory, we consider the Type-B scenario physically unlikely.

**Sensitivity of the results to the presence of prominent structures**

Our main conclusions are based on the comparison with the simulations of hydrodynamic turbulence. While this comparison is sensible, given that the Coma Cluster undergoes several mergers, there are other astrophysical effects that make the interpretation more difficult.

The central region of Coma shows quasi-linear, high-density structures of low-entropy gas likely stripped from merging subclusters (Figure 1b)[11,18] that span about 150 kpc. If we excise part of the image containing these structures from the analysis (Supplementary Figure 4a), the amplitude of fluctuations is reduced maximum by a factor of 1.2 on spatial scales ~ 50 – 110 kpc (coincide with the peak of the local bump at k ~ 0.009 – 0.02 kpc$^{-1}$). This brings the slope of the observed amplitude closer to the Kolmogorov model (Supplementary Figure 4d). Visual inspection of gas perturbations in the offset region did not reveal similar linear and extended structures.

There are two large cD galaxies, NGC4889, and NGC4874, surrounded by bright dense halos[11,30] in the central region. Deep XMM-Newton observations of the Coma Cluster identified X-ray structures associated with smaller, normal galaxies as well[32]. In addition to the normal galaxies, there is a population of ultradiffuse galaxies[33]. We excised from the image two large cD galaxies, identified normal galaxies, setting the radius of each region to 1.5 times the effective radius of each galaxy in the catalog, ultradiffuse galaxies taken from the deep Subaru surveys[33] (Supplementary Figure 4b,c) and repeated the analysis. The resulting amplitudes of fluctuations are consistent with the default amplitude within the statistical uncertainties in the central and offset regions (Supplementary Figure 4e,f). As expected, the exclusion of a large fraction of the data leads to an increase of Poisson errors. Finally, it was shown that compact galactic mini-coronae with the size ~ 3 kpc are present in the Coma image[34]. These structures are already excluded from the default analysis.

In addition to galaxies and stripped gas, Sanders et al.[11] claimed the presence of 50-kpc-long filamentary features likely related to the presence of magnetic fields. These features are very localized and unlikely to affect our volume-filling measurements. Planck SZ observations of Coma identified a shock front with Mach number ~ 2 at r ~ 900 – 1100 kpc in the west and south from the cluster center[35]. Since our offset region (up to ~ 550 kpc) does not reach the shock front, contribution of this shock to the observed amplitude is likely negligible. However, we cannot exclude the contribution of weaker shocks and sound waves generated by mergers and turbulence, which are likely present in such an unrelaxed, merging cluster. The order-of-magnitude estimates showed that sound waves do not produce density perturbations efficiently unless turbulent velocities are larger than ~ 450 km/s[10]. More accurate estimates require direct velocity measurements in Coma and an observational identification of shock fronts and sound waves, which are beyond the scope of this paper.

**Spherically-symmetric unperturbed models of X-ray surface brightness**

Position of the center of the Coma Cluster is ambiguous since the X-ray surface brightness is flat in the center. Our default choice of the cluster center is based on visual inspection of the contours of X-ray surface brightness in the central region. The radial profile of the surface brightness and the best-fitting model are shown in Supplementary Figure 5. We checked how the parameters of the model alter if the cluster center is shifted by ~ 20 kpc relative to the default choice. The core radius and β varied by no more than 5-6%. Such small uncertainties in the unperturbed model do not affect the results of the fluctuation analysis.

At larger distances from the cluster center, the cluster asymmetry and ellipticity become substantial. In fact, when dividing the image of the offset region by the spherically-symmetric model described above, the residual image shows a very strong spurious gradient of surface brightness in the approximately perpendicular to the radial direction. In order to account for this asymmetry, we obtain a model centered at a distance ~ 120 kpc from the cluster center (Figure 1) and only focus on the sector 0°-90° that includes our offset region. In this model, the radial direction is perpendicular to the gradient of surface brightness. The radial profile and model are shown in Supplementary Figure 5. We also tested other options, namely, we shift the center of the model by ~ 50 kpc relative to the default choice. In most cases, the differences in the best-fitting parameters are of the order of few % only. Such differences are negligible for the fluctuation analysis.

**Large-scale coherent structures and asphericity of the underlying model**

The amplitude of fluctuations in the offset region is slightly steeper than the Kolmogorov model, $k^{-1/3}$ (Supplementary Figure 2). This is due to the presence of large-scale coherent structures in the residual images of X-ray surface brightness fluctuations. To verify whether these large-scale structures affect the amplitude of fluctuations on small scales, in particular, on scales close to the mean free path, we re-did the analysis using the underlying β-model patched on large scales. Namely, we use $I_\beta S_\sigma[I_X/I_\beta]$, where $I_\beta$ is the spherically-symmetric β-model, $S_\sigma[\cdot]$ denotes Gaussian smoothing with the smoothing window σ, $I_X$ is the cluster X-ray surface brightness[36]. Supplementary Figure 6 shows the underlying patched β-models and the corresponding residual images of gas fluctuations. The corresponding amplitudes are shown in the right panels. For these experiments, we use smoothing windows with the width σ =200" and 150". By design, the patched β-models suppress the amplitude of fluctuations on large scales in both central and offset regions in Coma. If σ = 200" is used, then the spectra are affected on scales k < 0.005 arcsec$^{-1}$ (as expected). At large k, the amplitudes remain the same as in the case of a spherically symmetric, default β-model. If σ =150" is used instead, the effects are noticeable at slightly larger wavenumbers, k ~ 0.006 arcsec$^{-1}$. This experiment shows that the small-scale fluctuations are indeed present in the hot gas in Coma and their amplitude on small scales is not sensitive to the presence of large-scale structures and the choice of the underlying model.

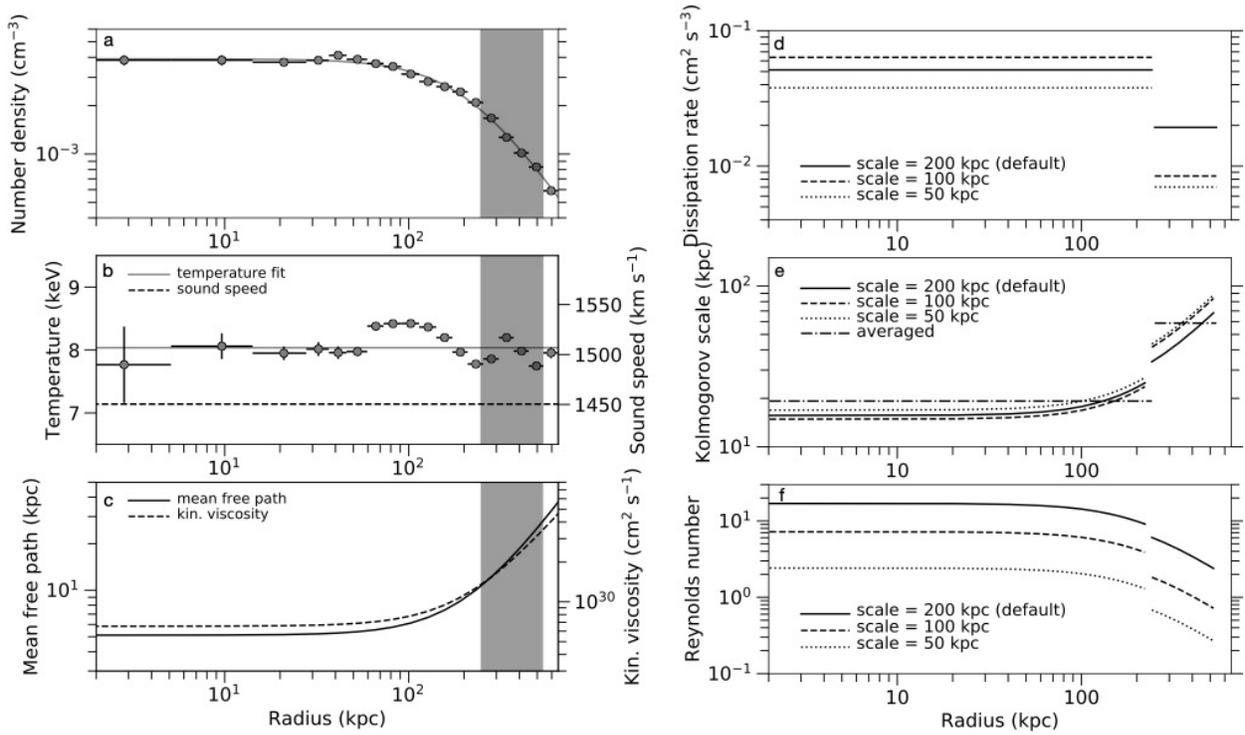

**Supplementary Figure 1 | Properties of the hot gas in the Coma Cluster.** (a) Measured deprojected number electron density (points with 1σ errorbars) and the best-fitting model (gray curve); (b) the electron temperature (points with 1σ errorbars) and the best-fitting value (gray line), the sound speed in the gas calculated from the best-fitting temperature (dashed line); (c) the mean free path of electrons and ions (solid curve), the kinematic gas viscosity (dashed curve); (d) the dissipation rate calculated from the observed amplitude of density fluctuations on spatial scales 200 (solid), 100 (dashed) and 50 (dotted) kpc; (e) the Kolmogorov microscale calculated on the same scales (200: solid, 100: dashed, 50: dotted) and averaged over scales and distance from the cluster center (dot-dashed); (f) the effective Reynolds number computed on the same spatial scales (200: solid, 100: dashed, 50: dotted). Gas characteristics shown on the panels c-f are calculated for plasma dominated by Coulomb collisions. We use the mean value of the amplitude of density fluctuations to obtain the values shown in panels d-f. Gray regions in panels a-c indicate the offset region in Coma.

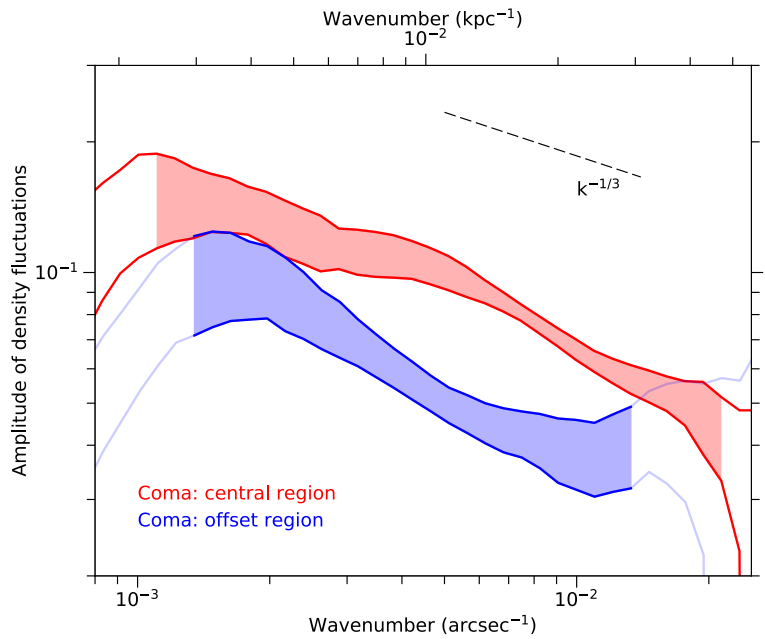

**Supplementary Figure 2 | Amplitude of density fluctuations vs wavenumber in the Coma Cluster.** The central region is shown in red, while the offset region in blue. Shaded regions indicate a range of scales where the amplitude is least affected by systematic uncertainties (e.g., removal of the unperturbed model, significant at low k) and by Poisson noise and the contribution of point sources (affecting mainly large k). The width of each region corresponds to 1σ statistical uncertainty. The dashed line shows the slope of the Kolmogorov model of turbulence, $k^{-1/3}$.

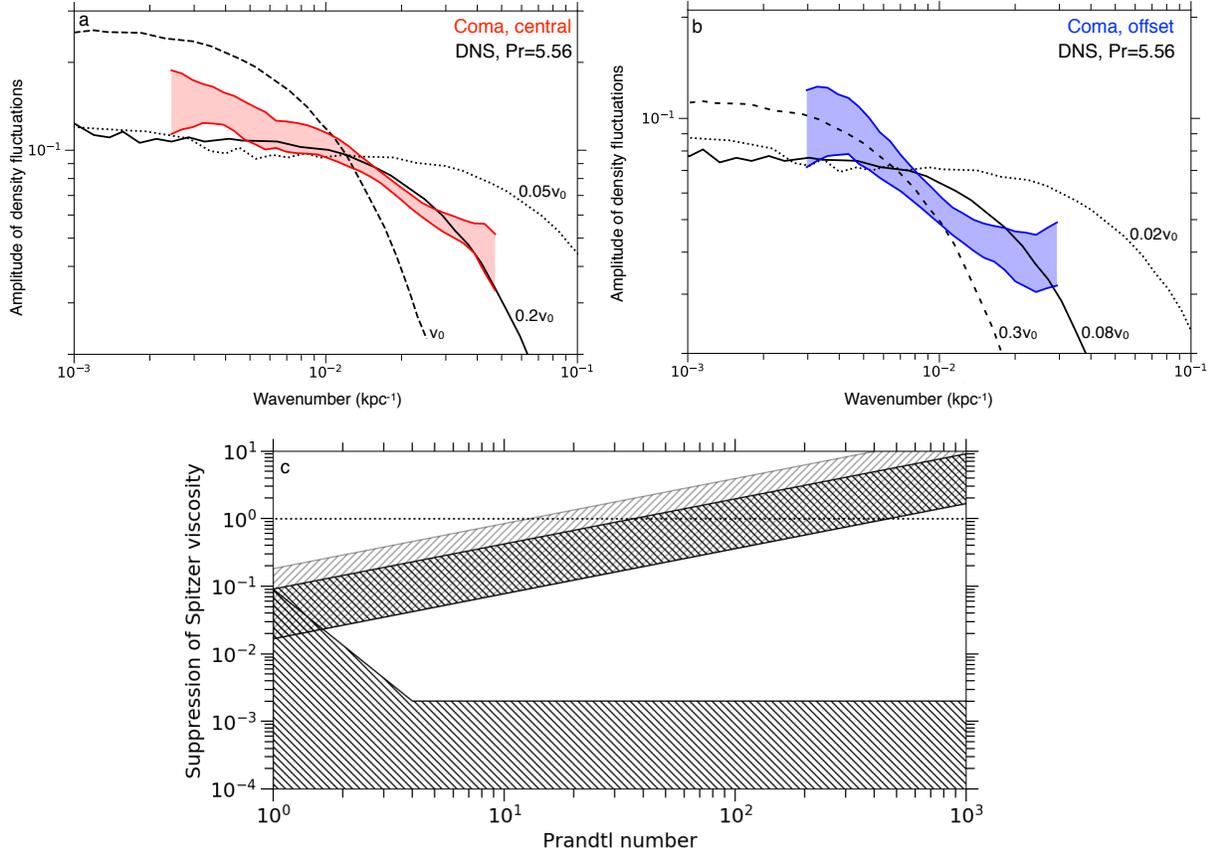

**Supplementary Figure 3 | Viscosity constraints for Pr > 1 in Coma.** (a) The amplitude of density fluctuations in the central region in Coma (red) in comparison with the passive-scalar amplitude from DNS[24,37,38,39] for Pr=5.56 (black curves). These curves are labeled according to the ratio of the adopted viscosity relative to the Spitzer value. Dashed, solid and dotted curves correspond to the ratio value 1, 0.2, and 0.05, respectively. (b) Same for the offset region. Dashed, solid and dotted curves correspond to the viscosity suppression by a factor of 0.3, 0.08, and 0.02, respectively. For Pr=5.56, the DNS spectra approximately reproduce the observations if the viscosity is suppressed by a factor of 0.2 (0.08) in the central (offset) region. (c) Suppression of gas viscosity in Coma relative to the Spitzer value vs Pr >=1. Black hatched regions show the estimated values of gas viscosity and Pr that approximately reproduce the observed data in both regions. Gray region accounts for systematic effects associated with linear, low-entropy structures in the central region and possible contribution from galaxies. The lower horizontal region with large viscosity suppression corresponds to the Type-A solution (see Methods), while the hatched diagonal region is the Type-B solution (see Methods). For guidance, the dotted line shows the case of full Spitzer viscosity.

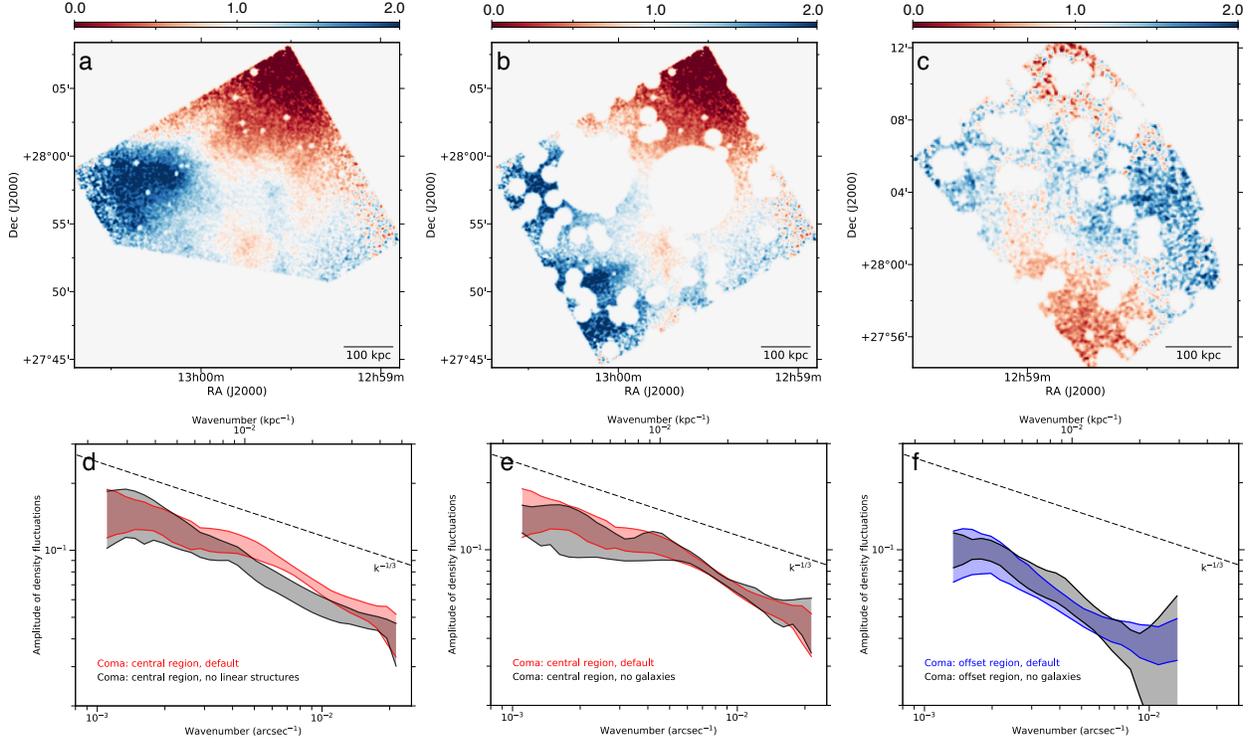

**Supplementary Figure 4 | Sensitivity of the amplitude of density fluctuations in Coma to prominent structures.** Top panels (a-c) show the residual images of Coma central (a-b) and offset (c) regions, from which we excluded quasi-linear structures of low-entropy gas (a), and galaxies (b-c). Bottom panels (d-f) show the default amplitude of density fluctuations (red or blue) and the modified ones (black) if specific structures are excluded from the analysis (see corresponding top panels). The exclusion of quasi-linear structures of low-entropy gas in the central region removes a bump on k~ 0.015 kpc$^{-1}$ bringing the amplitude closer to the Kolmogorov model (d). Exclusion of two large cD galaxies in the Coma center, other normal galaxies and ultradiffuse galaxies in the central and offset regions from the analysis slightly modifies the shape of the amplitude at specific k, however the overall slope of the spectrum remains consistent with the initial (default) amplitudes within the errors (e, f). The width of each amplitude (d-f) corresponds to 1σ statistical uncertainty. Dashed lines in the panels d-f show the slope of the Kolmogorov spectrum, k$^{-1/3}$.

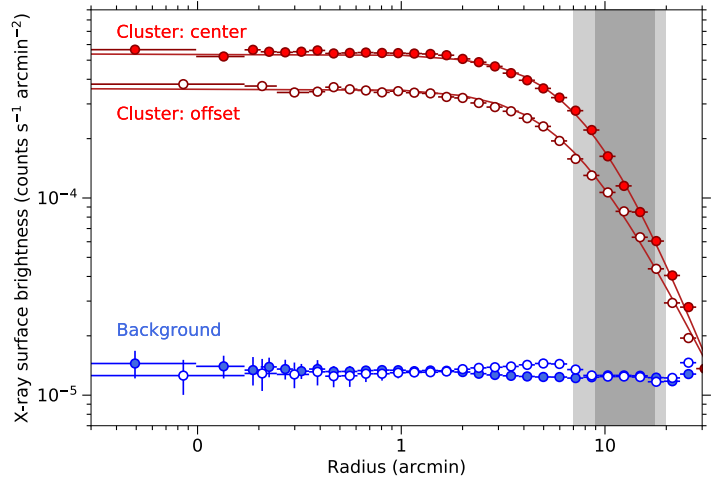

**Supplementary Figure 5 | Radial profiles of X-ray surface brightness in Coma.** Solid points show a spherically-symmetric radial profile centered at the cluster center, on RA=12h59m42.67 (J2000) and Dec=+27°56'40.9 (J2000) (Figure 1). The best-fitting β-model has a core radius 8.5' and β=0.6 (solid curve). Open points show the surface brightness profile in the 0°-90° sector centered on RA=12h59m21.629 (J2000) and Dec=+27°55'08.01 (J2000), which is ~ 120 kpc away from the central pointing (Figure 1). The corresponding best-fitting β-model has the core radius 6.8' and β=0.5. This model is used as an unperturbed model of surface brightness in the offset region since it better accounts for the cluster asymmetry and ellipticity at large distances from the cluster center. Red and blue points correspond to the cluster and background profiles, respectively. Error bars show 1σ statistical uncertainty. Vertical gray areas show the approximate radial position of the offset region.

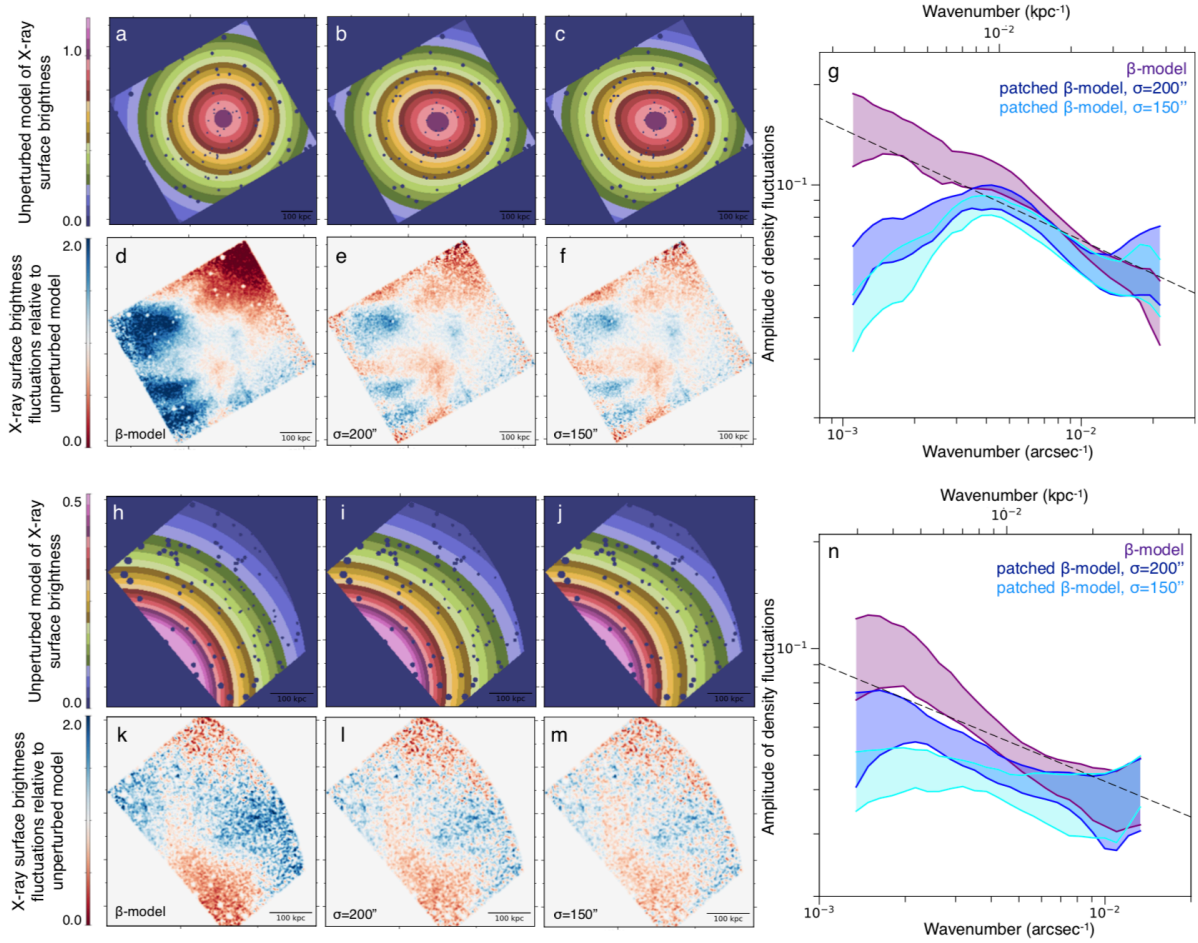

**Supplementary Figure 6 | Effects of large-scale coherent structures on the amplitude of density fluctuations in Coma.** Panels a-g (h-n) show the results of our experiments with patched β-models in the central (offset) region. In each set of figures, the top three panels show the underlying models: spherically symmetric β-model (left), patched β-model with σ = 200" (middle), patched β-model with σ = 150" (right). The corresponding residual images are shown in the bottom panels. The color scale of the residual images is the same in all panels. The right panels show the amplitude of density fluctuations calculated relative to these underlying models: purple/blue/cyan region corresponds to panel d/e/f in the central region (top panels) and k/l/m in the offset (bottom panels) region. By design, the amplitude is suppressed on spatial scales defined by the smoothing window size. The amplitude on small spatial scales, including scales close to the particle mean free path in the offset region, are not affected by the large-scale structures.